# Listener-Position and Orientation Dependency of Auditory Perception in an Enclosed Space: Elicitation of Salient Attributes

Bogdan Bacila and Hyunkook Lee

bogdan.bacila@hud.ac.uk, h.lee@hud.ac.uk

Applied Psychoacoustics Laboratory (APL), University of Huddersfield, Huddersfield, HD1 3DH,

United Kingdom

## Abstract

This paper presents a subjective study conducted on the perception of salient auditory attributes depending on the listener's position and head orientations in an enclosed space. Two elicitation experiments were carried out using the Repertory Grid Technique; in-situ and laboratory experiments aimed to identify perceptual attributes among ten different combinations of the listener's positions and head orientations in a concert hall. Results revealed that, between the in-situ and laboratory experiments, the listening positions and head orientations were clustered identically. Ten salient perceptual attributes were identified from the data obtained from the laboratory experiment. Whilst these included conventional attributes such as ASW (apparent source width) and LEV (listener envelopment), new attributes such as PRL (perceived reverb loudness), ARW (apparent reverb width) and Reverb Direction were identified, and they are hypothesised to be sub-attributes of LEV (listener envelopment). Timbral characteristics such as Reverb Brightness and Echo Brightness were also identified as salient attributes, which are considered to potentially contribute to the overall sound clarity.

## 1. INTRODUCTION

Research into the subjective perception of concert hall acoustics provides insightful knowledge that can be applied in multichannel audio and virtual acoustics. Virtual Reality (VR), Augmented Reality (AR) and Mixed Reality (XR), embraced under the term Extended Reality (XR), are rapidly growing areas of research and development, which are being increasingly used in various applications. To name a few, XR can be used in the entertainment industry as a new content delivery format, in which the user can interact with the content. Using virtual acoustic simulations of enclosed spaces, a virtual musical performance can be experienced in such a way that the listener can freely move around in a virtual or physical space. In such a scenario, it is necessary to dynamically render various acoustic features of the environment (e.g., early reflections and reverberation characteristics as well as source direction) according to the listener's position and head orientation. Highly accurate virtual acoustic rendering using conventional simulation tools tend to be computationally expensive and is not suitable for real-time audio rendering. Developments into perceptually optimized virtual acoustics can facilitate the real-time rendering of spatial acoustics in different environments. This however requires a solid understanding of psychoacoustical factors that influence the perception of various spatial attributes.

The complex nature of personal perception of sound phenomena raises a multitude of questions regarding the development of a commonly agreed lexicon for defining the spatial attributes perceived in room. Understanding about salient auditory attributes and their associated objective measures can provide an approach for measuring and reproducing different characteristics of a given acoustic environment.

Hawks and Douglas [1] and Blauert and Lindemann [2] observed the subjective perception of concert auditoria experience as being multi-dimensional, which required a new approach towards the research and methodologies used. Five dimensions were found most suitable to describe the experience: Reverberance, Balance and Blend, Intimacy, Definition and Brilliance [1]. Similar verbal descriptors were used by Barron [3] in his subjective study on British concert halls. A number of researchers (e.g. [4, 5]) have attempted the standardisation of particular auditory attributes. The term "Spatial Impression" was found to be related to the early lateral reflections as well as being strongly correlated with subjective preference. Morimoto and Posselt [6] also suggests that the effects of late reverberation is as salient as the early reflections in the subjective perception of a concert hall. Studies after 1990s (e.g. [7, 8, 9] ) tend to agree that there are two dimensions which can define the overall spatial impression of a concert hall: ASW (Apparent Source Width) and LEV (Listener Envelopment). To this day, this is still the most widely used paradigm for describing concert halls.

Rumsey [10] claimed that the simple paradigm of ASW and LEV is insufficient for accurately describing different perceived attributes in sound recording and reproduction. He proposed a "Scene-Based Paradigm", where a number of low-level perceptual attributes are separated into more ecologically valid categories such as source descriptors, environment descriptors and global scene attributes. However, as Rumsey's study was focused mainly on reproduced audio it is important to assess the validity of this classification in the perception of concert hall acoustics, through additional elicitation tests.

Lokki et al. [11] researched into the perceptual attributes in concert halls. A virtual orchestra was used in three concert halls for stimuli creation and multichannel 3D reproduction was used for the experiments. Sensory profiles were created by eliciting perceptual attributes from the three different concert halls. Analysis revealed that loudness and distance had a strong perceptual effect and reverberance could be associated with either the size of the space or the enveloping reverberance. Studies by Lokki et al. [12] [13] and Kuusinen et al. [14] take a more in-depth look into the relationship between the elicited perceptual attributes, objective measurements and ultimately preference.

Studies such as [15][16] summarized a large amount of information regarding lexicon used for describing sound. In [15] a sound wheel was developed, providing a common and structured lexicon to be used in describing reproduced sound. However, limited emphasis was put onto the identification of spatial attributes as the main focus was placed on timbral characteristics. Zacharov and Pedersen [16] expanded further on the sound wheel by studying the spatial sound attributes. Semantic data mining and clustering revealed 50 attributes which can be used to describe different aspects of spatial impression.

Whilst the aforementioned research is valuable in recognising a wide range of perceptual descriptors for different concert halls, there is a lack of understanding in the perception of auditory attributes depending on the listener position and head orientation. Research conducted by Mason et al. [17] highlighted the perceptual effects caused by the head-rotation-dependent variations in inter-aural cross correlation coefficient (IACC). In their study, two experiments were carried out testing how variations in IACC affect different perceptual attributes depending on different head movements of the listener's head. Synthesized IACC variations were tested in the first experiment and variations caused by different loudspeaker configurations were used during the second experiment. Results showed that the width and depth of the reverberant environment, as well as the perceived width and distance of the source, were strongly influenced by the variations in IACC. In a later study, Lee [18] investigated into the perception of ASW and LEV at different source-listener distances in a reverberant concert hall. Perceived ASW was found to statistically decrease almost linearly as the distance from the source was doubled. Similarly, LEV was found to decrease with doubling the distance, but with a lower magnitude. However, only three listening positions were tested, all on the central axis of the room and all facing towards the source.

From the above background, the present study aims to determine what kinds of low-level auditory attributes can be perceived depending on the listener's position and head orientation in a concert hall. For this, both in-situ and laboratory experiments were conducted to elicit subjective descriptions of perceived attributes. As both experimental environments presented inherent limitations, the experiments were designed to complement each other for a comprehensive understanding of the spatial impression. The original repertory grid technique methodology was extended with an audibility-based filtering during the laboratory experiment in order to identify the most salient perceptual attributes.

It is expected that a more fundamental understanding about the perception of spatial impression in a multi-position and multi-head-orientation context can greatly improve the VR/AR experience with faster, more precise, perception-based acoustic models. Development of relevant rating scales for perceptual spatial impression attributes is an important initial step towards more in-depth research into the perception of auditory attributes in a 6DoF context.

## 2. EXPERIMENTAL DESIGN

Research into the perception of auditory attributes usually requires the adaptation and implementation of methodologies and techniques used in other fields such as food evaluation and psychology. This section presents the elicitation methodology and experimental procedures employed during this study.

### 2.1 General Methodology

### 2.1.1 Repertory Grid Technique (RGT)

Several different descriptive elicitation methods such as quantitative descriptive analysis (QDA) [19], Spectrum Descriptive Analysis [20], Flash Profile [21], Individual Vocabulary Profiling (IVP) [22] and Repertory Grid Technique (RGT) [23] have been used in the fields of concert hall acoustics and spatial audio. For the present study, RGT was chosen for the following reasons; It is an individual vocabulary method, similar to Free Choice Profiling, but uses a more structured approach for construct generation. Opposed to a consensus vocabulary method such as QDA, where more experienced subjects can dominate the conversation, an individual vocabulary enables all subjects to provide more input. For the present research it was important to encourage the generation of audio attributes from all subjects, thus an individual vocabulary method was desired. RGT requires the assessment of a particular experience both in its similarity and its dissimilarity to other experiences, therefore facilitating the elicitation of personal constructs. RGT was previously used by Berg and Rumsey [24] as an attribute elicitation method for spatial audio reproduction. Similarly, McArthur et al. [25] used this method in the process of elicitation and analysis of perceived auditory distance in cinematic VR.

RGT is structured in two separate stages: construct elicitation and construct rating. In the construct elicitation stage, triads of stimuli are presented to the test subjects, who are asked to describe in which way two of the stimuli are more alike and how they differ from the third one. A new triad is presented and assessed until there are no more constructs to provide. This step generates a set of opposing attributes (e.g., quiet-loud, close-far), named "bipolar constructs".

In the following construct rating stage, each of the elicited bipolar constructs is used as a 5-point grading scale ranging from 1 to 5, with 1 representing one pole of the elicited construct and 5 representing the other pole. Each stimulus is then graded accordingly on the newly formed attribute scales. The grading stage of RGT ensures that the elicited personal constructs are rated at each listening position. Thus, constructs rated similarly across the entire data set might suggest an underlying perceptual similarity even though different terminologies were used to describe the attributes. This is particularly useful in the present research for observing how certain lexicon is used in combination with the perceived spatial impression attributes and will aid in the generation of new attribute names.

### 2.1.2 In-situ and laboratory experiments

The present study conducted both in-situ and laboratory experiments to complement each other. The in-situ test was considered to be an important first step from the viewpoint of ecological validity; being physically in the space would allow the subject to perceive auditory attributes in the most natural and accurate way. Although the listener's auditory perception may be biased by other senses such as vision, data collected from the in-situ investigation would reflect real life experience. However, a limitation of the in-situ RGT experiment with a multi-position comparison task was identified; it was not easy for the subject to compare different test conditions since the subjects had to physically move between different positions whilst remembering the auditory experience of the previous condition. This may have made it difficult for the subjects to detect subtle differences.

Considering this limitation, a laboratory RGT experiment was also conducted using binaurally auralised stimuli reproduced over headphones. This allowed instant switching between the stimuli, thus enabling the subjects to compare different conditions more effectively and efficiently. The lab experiment also allowed the present authors to control the experimental environment more. However, it is important to verify the ecological validity of the results obtained in the virtual environment against those from the in-situ investigation, which will be discussed in Section 4.3.

Furthermore, this study extended the traditional RGT with an additional audibility grading phase in order to identify most salient attributes. As mentioned earlier, RGT provides a well-structured framework for the elicitation of personal constructs. However, it does not generate any information about which attributes are perceptually more important. The second stage of the RGT grades each attribute for each stimulus, however, this grading is only used for assessing the level of similarity between different constructs or stimuli. Therefore, the relative difference between different stimuli in perceptual strength is not known from the RGT result. In the context of the present study, it was also of interest to see which of the elicited attributes have stronger audible differences among different position/orientation conditions. Therefore, after the lab RGT test, an audibility test was carried out to compare all of the stimuli conditions for each of the attributes elicited from each subject. The subjects were asked to rate the perceived magnitude of the difference between all of the conditions, for each of their own elicited bipolar constructs. This allowed for further filtering of the constructs based on their reported audibility rating.

## 2.2 In-Situ Experiment

### 2.2.1 Physical setup

The experiment was carried on in the University of Huddersfield's St Paul's concert hall (Fig.1, RT = 2.1s; 16m (W) x 30m (L) x 13m (H)). A Genelec 8040A loudspeaker was placed in the centre of the stage and used for exciting the room. An anechoic male speech taken from the Bang and Olufsen's Archimedes database [26] was used as a test signal, due to its broadband frequency nature and controlled ratio of transient and sustained sound. Four positions from the concert hall, with different orientations were tested. A total of ten listening position and head orientation conditions were assessed: four facing towards the source, four facing 90° from the source and two facing away from the source (Figure 1). P1_0°, P1_90° and P1_180° were situated on the room's centre line, at 2m from the loudspeaker. Similarly, P2_0°, P2_90° and P2_180° were situated on the same line at 8m away from the source. P3_0°, P3_90°, P4_0° and P4_90° were situated laterally, 4 metres away from the centre line, at the same distances from the source as the previous groups. For the aforementioned positions, two head orientations, one facing towards the source and one facing 90° were tested. The measured loudness for each of the positions is presented in figure 1.

It is to be noted here that the listening positions are situated between the stage and the seating area - situated approximately ten metres away from the stage - for a number of reasons: the underlaying motivation of this research relies in finding salient perceptual characteristics of a space, to be used in an XR scenario. Consequently, a space where one can freely move from one position to another is desirable for future experiments to be carried out in a VR scenario. It was also decided that other rooms, such as lecture theatres or classrooms will be measured and tested therefore changes caused by different heights in the seating area were undesirable for the present research.

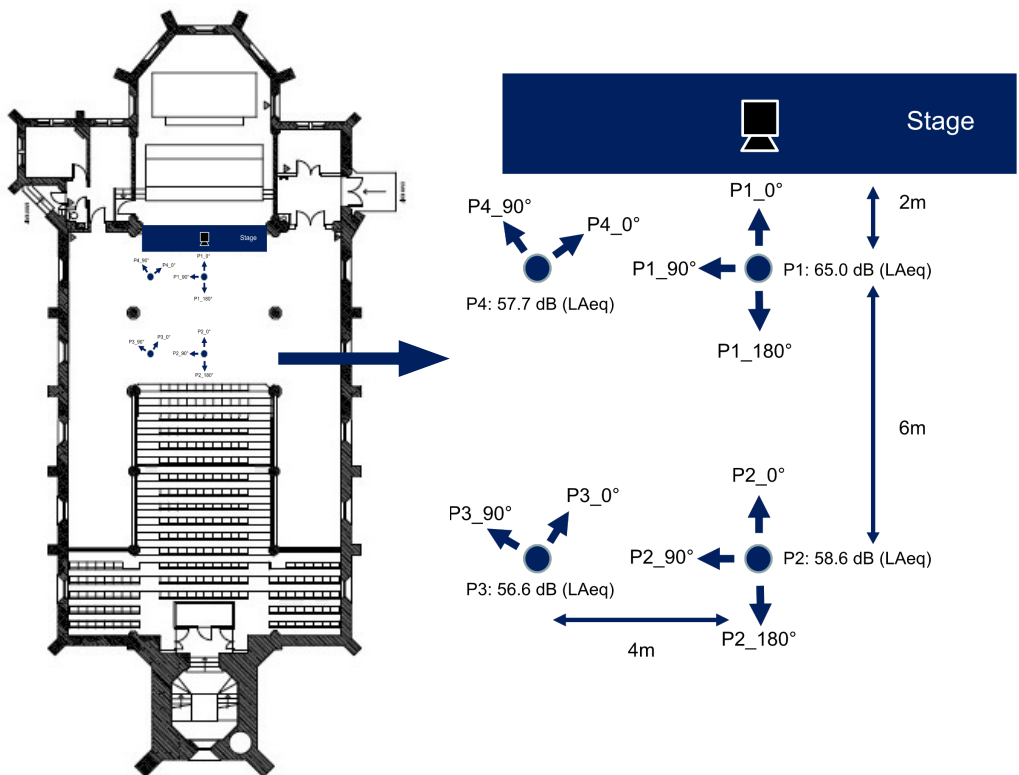

Figure 1. The top view of the St. Paul's concert hall and the listener positions and orientations used for the elicitation experiment.

### 2.2.2. Subjects

A focus group consisting of six subjects took part in the in-situ experiment. They were postgraduate students and academic staff from the Applied Psychoacoustics Lab (APL) of the University of Huddersfield, with their ages ranging from 23 to 42. All of them reported to have normal hearing. The group was carefully selected for the following reasons. They were highly familiar with the acoustics of the space since they had accessed the space regularly for spatial recording and critical listening sessions for at least one year previously. Furthermore, they had an extensive experience in various types of critical listening test, and also had an intermediate or advanced level of knowledge about auditory perception and concert hall acoustics. Since the nature of the elicitation task required a critical listening skill and the ability to describe perceived auditory differences precisely, the use of the small group of such experienced and trained subjects was considered to be more beneficial than using a larger number of naïve subjects, who may not be able to provide sufficient and reliable data.

### 2.2.3. Test protocol

The test was split into two sessions for each subject; elicitation and construct rating sessions in sequence. The first session took approximately 40 minutes and the second one approximately 20 minutes. During the first session, the subjects were encouraged to take a short break after 20 minutes in order to avoid listening fatigue. Prior to the test, the subjects were allowed to spend some time walking around in the space while listening to the speech for familiarisation.

In the elicitation stage, subjects were asked to compare three conditions at a time, by physically moving from one to another. The triads were provided on a response sheet where they also recorded the elicited terms. While assessing the spatial impression attributes they were asked to think about how two of the conditions are similar and how do they differ from the third one. This method of comparison ensured that elicited result is a bipolar construct (e.g. wide – narrow source). Because of the long distance between the positions it was decided to provide the triads in a consecutive order. ("P1_0°, P1_90°, P1_180°"- triad one, "P1_90°, P1_180°, P2_0°"- triad two, etc.). All conditions were tested during the experiments and the elicitation stage ended when there were no more constructs to elicit. All participants had to asses at least 10 triads in order to experience all conditions. After the 10 mandatory triads they were encouraged to test any other personal combinations of stimuli until all the attributes have been elicited.

After finishing the elicitation process, the resulting bipolar constructs were arranged in a grid, with the columns representing the positions under test and one of the bipolar constructs on either side. The participants were asked to check the grid and the constructs for consistency with their own vocabulary, similarly to the experiment described by Berg et al. [24]. In the rating process subjects had the task of assessing each condition individually, based on their own provided attributes. The conditions were randomized in order to avoid any bias and the order noted down on the response sheet. Participants walked to each position and graded each listening condition on a scale from 1 to 5, with 1 corresponding to the left pole and 5 to the right pole. (Appendix 1-2)

### 2.3 Laboratory Experiments

### 2.3.1 Stimuli creation

Binaural stimuli were created by convolving the same anechoic speech sample previously used in the in-situ test with binaural room impulse responses (BRIRs) captured at the same source and receiver positions as well as head orientation angles used in the first experiment. The BRIRs were acquired using the exponential sine sweep technique [27] in the HAART software [28], with a Neumann KU100 dummy head microphone and a Genelec

8040A loudspeaker. These BRIRs are part of an open-access database[1], and more details about it are provided in [29]. The playback level of the binaurally synthesised stimulus for each position was adjusted to match the perceived loudness of the corresponding stimulus used in the previous in-situ test.

The experiment allowed a small amount of head rotation to help the subjects resolve a potential front-back confusion as well as to provide a more natural listening experience. This required the binaural stimuli to be synthesised dynamically according to head tracking. Therefore, for each of the stimuli, five BRIRs ranging between -7.2° and 7.2° with the intervals of 3.6° were used. A real time interpolation was achieved between each adjacent BRIR by the VBAP method [30]. The head tracker used was based on the method developed by Romanov et al. [31], which was adapted using Bluetooth for a wireless transmission of the position information. The head rotation was limited by visual indicators that guided the subject to reposition their head in the centre position if they exceeded the +/- 7.2° limit.

### 2.3.2. Physical setup

Both the elicitation and audibility tests took place in an ITU-R BS.1116 [32]-compliant listening room at the APL. Graphical user interfaces (GUIs) for the tests were created in Max 7 and the stimuli were presented through a pair of AKG K702 headphones via a Merging Technologies Anubis interface. The stimuli were convolved with an inverse filter for the equalisation of the headphones, which was created from the average of ten impulse responses measured using the KU100 dummy head and the headphones.

A picture of the scene in the direction of the head orientation at each listening position was also shown on a separate screen (Fig 2). Visual information of the venue was part of the multimodal perception in the in-situ experiment. Therefore, it was aimed to create an auditory-visual test environment that is as similar as the real-life one. Using a head-mount display for a 360° virtual reality presentation was initially considered. However, this was deemed to be unnecessary as the subjects virtually faced forward during listening. Moreover, wearing heavy equipment over head during the test might increase listener's fatigue level and make the response process (typing on a keyboard) inconvenient.

---

[1] https://zenodo.org/record/2641166#.XxmCL_hKhIY

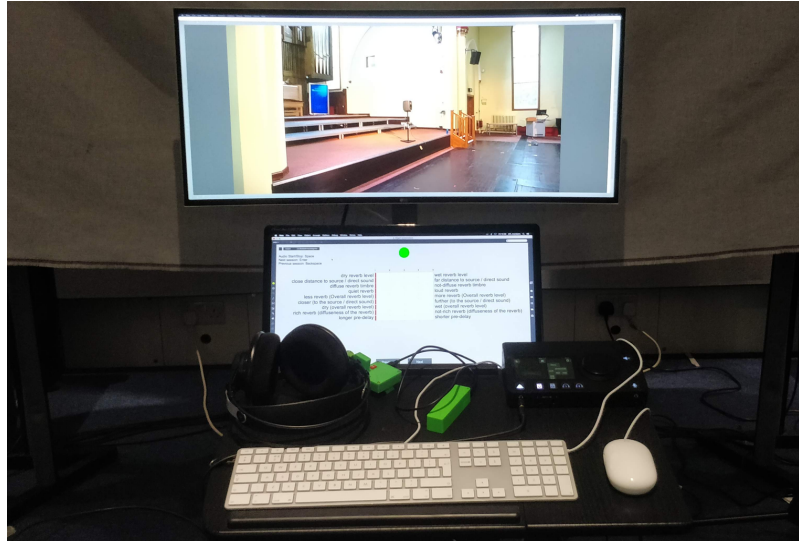

Figure 2. Lab elicitation setup

### 2.3.3. Subjects

Eight participants took part in the elicitation and audibility grading tests. Four of them took part in the in-situ experiment. The other four were final year students from the Music Technology and Audio Systems course of the University of Huddersfield. They all had previous experiences in spatial audio listening tests, and also were familiar with the acoustics of the venue from their previous trainings in sound recording conducted in the venue. They also had basic understandings about conventional spatial descriptors such as ASW and LEV.

### 2.3.4. Test protocol

*RGT Elicitation test*

The RGT procedure used for the elicitation test was identical to the one used in the in-situ test, apart from the fact that the subjects could instantly switch between the stimuli for each triad and provided their responses in a text box on the test GUI (Fig. 3(a)). Before starting the test, each subject completed a familiarisation stage where they were presented with the GUI and listened to all stimuli. As in the in-situ test, the listening test consisted of two separate stages: elicitation and construct rating for each stimulus. Subjects were required to take a short break after approx. 20 minutes in order to avoid listening fatigue.

*Audibility test*

An additional audibility test was conducted for the reason described in Section 2.1.2. In each trial of the test, each subject was asked to listen to and compare all of the stimuli provided on the GUI (Fig. 3(b)) and rate the perceived magnitude of the difference among all of the conditions for each of their own bipolar construct of descriptive terms resulted from the RGT elicitation. A continuous scale with the internal scale from 0 to 100 and semantic anchors dividing the scale into four sections were used: just audible (0-25), slightly audible (25-50), audible (50-75) and very audible (75-100). The order of bipolar constructs to rate in each trial was randomised. After completing the audibility test, subjects were asked to verbally group some of their elicited attributes in terms of similarity. This additional step was used during the analysis along with attribute clustering for obtaining a better understanding of different verbalisations used for the same perceptual effect.

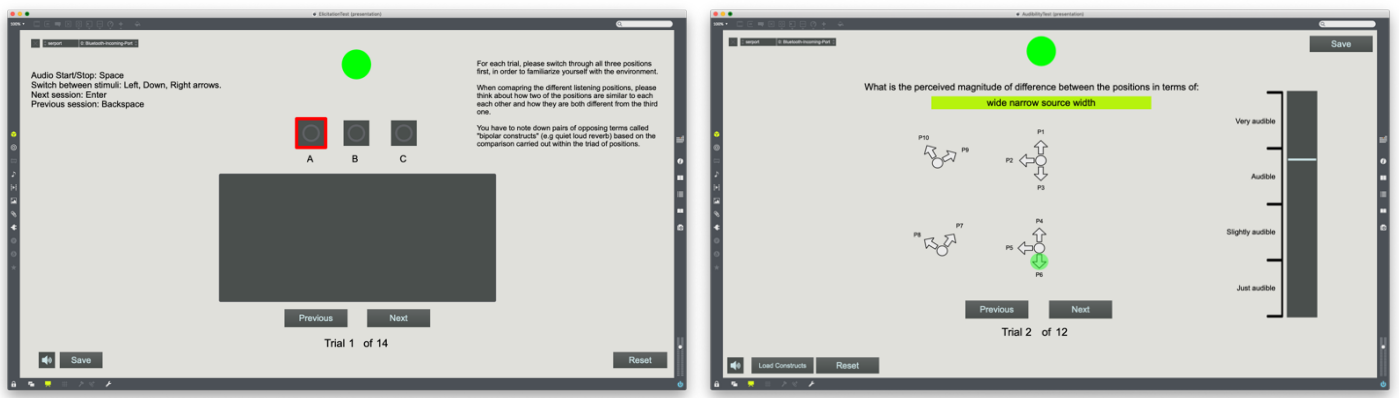

Figure 3. Graphical user interfaces used for the lab experiments. (a) Elicitation test (b) Audibility test

## 3. RESULTS

This section analyses the data collected from the in-situ and lab experiments described above. Before attempting any analysis of the repertory grids, the elicited constructs were analysed using a verbal protocol analysis (VPA) method as in Berg and Rumsey [24]. This delivered useful insights upon the semantic classification of the results. 56 and 98 constructs were elicited from the in-situ and lab experiments, respectively. An overview of the different analysis methods is presented in Figure 4 and a detailed description of each step is reported in this section.

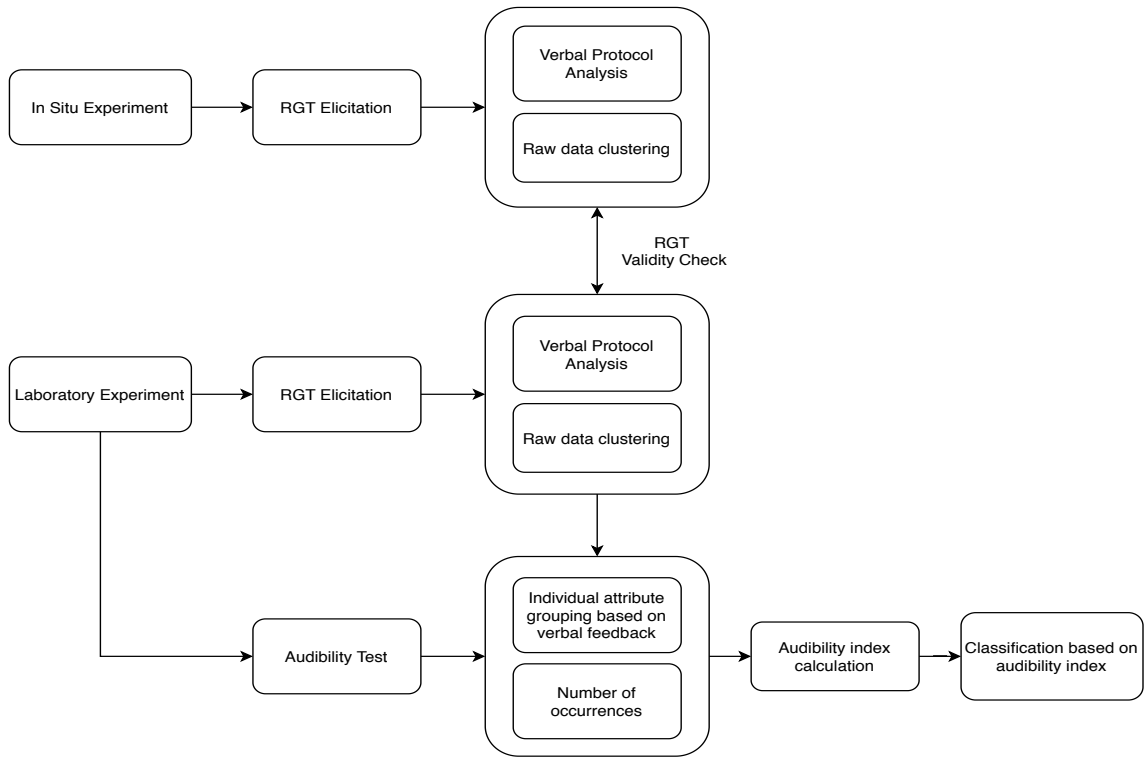

Figure 4. Analysis workflow

## 3.1 Verbal Protocol Analysis (VPA)

VPA was devised and used by Samoylenko et al. [33] as a method for distributing different verbalisations used in comparing musical timbres based on their semantic meaning. In the context of the present research, the subjective nature of the elicited data requires this type of analysis in order to observe different characteristics such as the sensory modality. The analysis of the verbalisations is carried out on three different levels: their logical sense, stimulus-relatedness and semantic aspects. This method of analysis was previously used by Berg and Rumsey [24] and McArthur et al. [34] for evaluating constructs resulted from RGT elicitations. VPA was also used by Zacharov and Koivuniemi [35] in combination with a QDA elicitation methodology.

The constructs elicited during this study were analysed based on the VPA Level 3 Features ("Semantic aspects of verbal units") as presented in Fig. 5. As in Berg et al. [24], VPA was used for observing the ratio between descriptive and affective features with the final goal of filtering only descriptive attributes for clustering. The constructs were initially classified as either "descriptive" (dfe) or "attitudinal" (afe) features. The descriptive features were subsequently categorised as either "unimodal" (umd – attributes referring only to the auditory modality) or "polymodal" (pmd – attributes referring to multiple sensory modalities). The attitudinal features were also classified as either "emotional-evaluative" (emv – describing particular emotions about the stimulus) or descriptors referring to the naturalness of the sound (ntl).

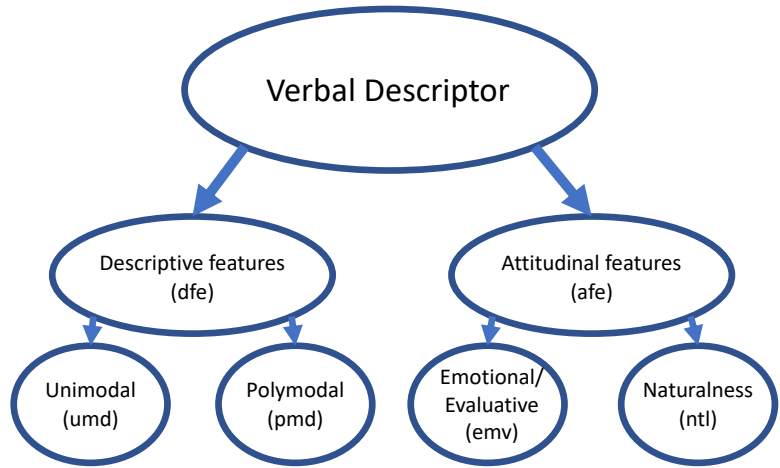

Figure 5. VPA Level 3 Features (after Samoylenko et. al [33])

As can be observed from Table 1, 55 out of 56 constructs (98.2%) were considered descriptive features for the in-situ test and 96 out of 98 (98%) for the lab elicitation test. The participants' previous experiences with spatial acoustic evaluation and the assessed space seem to be reflected in the low number of attitudinal features elicited (good/bad spatial quality, pleasant/unpleasant listener envelopment, close/distant claustrophobia), suggesting that participants were mainly focused on describing the spatial and timbral aspects of the sound events rather than subjective and emotional aspects. Unlike Berg's study [24], it was decided not to discard the attitudinal features from analysis in the present research to offer a better understanding of any emotional values associated with particular clusters of descriptors.

Table 1. Distribution of attributes in different categories after VPA

(a) In-Situ Experiment

| Feature | Number | % | dfe/afe | Number | % |
|---|---|---|---|---|---|
| Descriptive (dfe) | 55 | 98.2 | umd | 21 | 37.5 |
|  |  |  | pmd | 34 | 60.7 |
| Attitudinal (afe) | 1 | 1.8 | emv | 1 | 1.8 |
|  |  |  | ntl | 0 | 0 |

(b) Lab Experiment

| Feature | Number | % | dfe/afe | Number | % |
|---|---|---|---|---|---|
| Descriptive (dfe) | 96 | 98 | umd | 49 | 50 |
|  |  |  | pmd | 47 | 48 |
| Attitudinal (afe) | 2 | 2 | emv | 2 | 2 |
|  |  |  | ntl | 0 | 0 |

## 3.2 Cluster Analysis

Analysis of personal constructs from a purely semantic point of view might be difficult due to the underlaying subjective nature of the elicitation. Subjects describing the same perceptual effect may use different verbalisations. Therefore, the second stage of RGT facilitated the use of cluster analysis through the individual grading of each attribute. The results of clustering are presented as a dendrogram, in which the closer two constructs were (i.e., a smaller inter-construct distance), the more similarly they were rated. Clustering was achieved using the R statistical analysis software together with the OpenRepGrid package[36]. The inter-construct distance was calculated using the Manhattan metrics and the clustering was accomplished by Ward's method [37].

The ratings of the constructs for each position gathered from the in-situ and lab RGT experiments are presented along with the dendrograms in Appendices 1 and 2, respectively. As mentioned above, the audibility rating was used in the lab experiment for providing a more concise set of attributes to be used in further studies. Thus, more emphasis will be placed on the analysis and discussion of attribute clusters generated during the lab experiment and their saliency level based on the audibility ratings.

### 3.2.1 In-Situ experiment results

Five clusters were found at an inter-construct distance of 40 and ten clusters at a distance of 20. An overview of the clusters resulted from the analysis revealed a number of similar interactions between constructs across different clusters. Table 2 presents the main categories extracted from the cluster analysis. Whilst the observed clusters do not show a definite delimitation between the different attribute categories, certain aspects of spatial impression can be extracted from each cluster showing a possible perceptual or physical property.

Different verbalisations possibly suggesting similar effects to ASW and LEV were observed. For instance, constructs such as "source size" (No. 5 in Appendix 1) and "horizontal spread" (No. 1 in Appendix 1) might indicate a perceived attribute similar to "source width" (No. 3 and 4 in Appendix 1). As such, it is believed that even without a definitive attribute name, the perceptual effect can be linked to the more classical ASW. Attributes such as "reverb width" (No. 26) and "environmental width" (No. 23 and 25) might be associated with "envelopment" (No. 24), based on [10]. Again, multiple verbalisations could convey the same perceived spatial attribute of being "surrounded by reverb" (No. 28), which is in line with the classical LEV perceptual definition. The advantage of using RGT and cluster analysis can be well observed in cluster 3.3 (Appendix 1). A small cluster formed by no. 29, 30 and 31 refer to physical properties of the reverberation (time, loudness, D/R ratio). These are clustered with the aforementioned perceptual sense of being surrounded by reverberation, thus suggesting that these physical properties might be associated with the perception of "envelopment". New attributes describing the direction of the early reflections and reverberation relative to the central axis were elicited (No. 9,18,54). It is considered that these attributes are results of testing multiple listening position and multiple head orientations.

It is worth noting here that the scope of the present study was to define salient perceptual attributes for different listener positions and head orientations. This study is the first step towards the establishment of a cause and effect relationship between low-level and high-level attributes, which will be investigated in a future quantitative study.

Table 2. In-Situ main construct categories based on clustering

| Cluster | Attribute |
| --- | --- |
| 1.1 | Source width/spread |
| 1.2 | Reverb/Echo direction (Front/Side) |
| 2 | Echo perception and direction/Source width/Envelopment |
| 3.1 | Envelopment |
| 3.2 | Environment size/Source distance/Reverb level |
| 3.3 | Reverb spread/Envelopment/Reverb level |
| 4.1 | Source/reverb focus/diffuseness/clarity |
| 4.2 | Source Level |
| 4.3 | Reverb Spread/Source clarity/Environment depth |
| 5 | Envelopment/Reverb spread/Reverb direction |

Additionally, cluster analysis could be performed on the different variables, in this case the listening positions and head orientations used as stimuli. Four clusters could be observed (Appendix 1) showing similarities between different listening positions. Cluster 1 brings together P1_0°, P1_90° and P1_180°, the closest positions to the sound source. Custer 2.1 showed a similarity between P2_90°, P3_90° and P4_90°. This is particularly interesting as these positions are all facing 90º from the source. P3_0° and P4_0° can be observed in Cluster 2.2 and they represent the two positions facing towards the source, but from and off-centre listening position. Ultimately, P2_0° and P2_180° (Cluster 2.3) represent two of the positions on the centre line, one facing towards the source and one facing away from the source. The influence of listening position and head rotation will be discussed further in Section 4.2.

### 3.2.2. Lab experiment result: raw data clustering

Appendix 2 presents the elicited constructs along with their individual ratings for each position. Dendrograms resulted from cluster analysis are presented for both constructs and positions, showing the general interaction between different elicited terms.

Seven clusters were found at an inter-cluster distance of 40, and twelve clusters at a distance of 20. It can be seen that the elicited attributes generally fall into a number of recurring categories, presented in Table 3. An initial observation showed an increased number of constructs describing timbral differences in the virtual test than in the in-situ test. It is believed that the ability to instantly switch between the stimuli facilitated the perception of such timbral characteristics, which describe either the source, the perceived reverberation, or the room itself. Similar attributes as in the in-situ test could be extracted; For instance, in cluster 2 (Appendix 2) elements describing the direction of the reverb and echoes are tightly clustered. Likewise, similar attributes can be observed in other clusters as well (No. 12, 73, 74, 80, 86). Among some of the clusters it can be noticed again that different verbalisations might refer to the same perceptual effect. In Cluster 3.1, elements describing the direct sound can be recognised (No. 20, 23, 24, 25, 26, 27). Whether they refer to the perceived loudness of the source, intelligibility or spread, this seems to suggest that there might be a cause-effect relationship between the loudness of the source and its intelligibility or spread. A similar effect is seen in cluster 4.1 where the interaction between the source sharpness (No. 34), intelligibility (No. 35, 36) and the perceived distance to source (No. 38, 39) is apparent within a cluster focused tightly around the direct sound. Interestingly, Cluster 4.2 is also well focused around the reverberation. The elements describing the perception of the overall reverb loudness (No. 41, 42, 43, 44, 45) interact with the perceived reverb spread (No. 47) but also with the source width (No. 46). It is therefore apparent that there is a clear delimitation in the perception of source-related or room related attributes, which is also in-line with Rumsey's "Scene-Based Paradigm" [10].

Position clustering was carried out for the lab experiment results as well. As seen in Appendix 2, four clusters could be observed bringing similarly rated positions together. In Cluster 1 it can be seen that P1_0°, P1_90° and P_180° are similarly rated by participants. Cluster 2.1 brings together P2_0° and P2_180°, whilst in Cluster 2.2 positions P2_90°, P3_90° and P4_90° are grouped together. Finally, P3_0° and P4_0° are included in Cluster 2.3. These results are in complete agreement with the ones from the in-situ test, which seems to support the ecological validity of the laboratory experiment.

Table 3. Lab experiment main construct categories based on clustering

| Cluster | Attribute |
|---------|-----------|
| 1 | Reverb spread, Envelopment, Reverb direction |
| 2 | Reverb/Echo direction |
| 3.1 | Direct sound (level, clarity, timbre) |
| 3.2 | Source width, Early reflections, Reverb spread |
| 4.1 | Perceived distance to source/ source clarity, source spread |
| 4.2 | Reverb level |
| 5.1 | Envelopment, Reverb timbre, Room size perception |
| 5.2 | Source level, General timbre, Envelopment |
| 5.3 | Early reflections direction, perceived source distance/level, room timbre |
| 6.1 | Perceived reverb loudness, reverb direction |
| 6.2 | Envelopment, source width, source clarity |
| 7 | Echo direction, source spread, source loudness, reverb spread, |

### 3.2.3 Lab experiment result: Saliency based on audibility testing

The saliency for each attribute was determined from the audibility test, where subjects were required to note how audible the differences among all stimuli were, for each of their own elicited constructs. Figure 6 presents the individual results generated by the audibility testing.

A number of steps were employed in calculating an "audibility index", based on which a final classification could be outlined. The first step was grouping of similar attributes for each subject individually. The grouping was carried out based on both individual clusters and verbal feedback received from each participant regarding the similarity of their attributes. Once all the similar attributes were grouped, the most representative name was assigned, and their audibility scores were averaged. For the second step, the number of occurrences for each attribute among all the subjects was counted, resulting in a further reduction in the number of attributes. The cluster analysis was also used as a basis for grouping the most similar perceptual attributes. The audibility scores were averaged for each attribute group and an audibility index was calculated as follows: a percentage of the number of occurrences as well as the percentage of the audibility scores were calculated. An equally weighted average of the two was obtained, which was defined as an audibility index. The inclusion of the number of occurrences as well as this newly calculated value allowed for an easier observation and classification of the attributes based on the audibility index. Ten attributes which are presented in table 4 were observed to have an audibility index higher than 50.

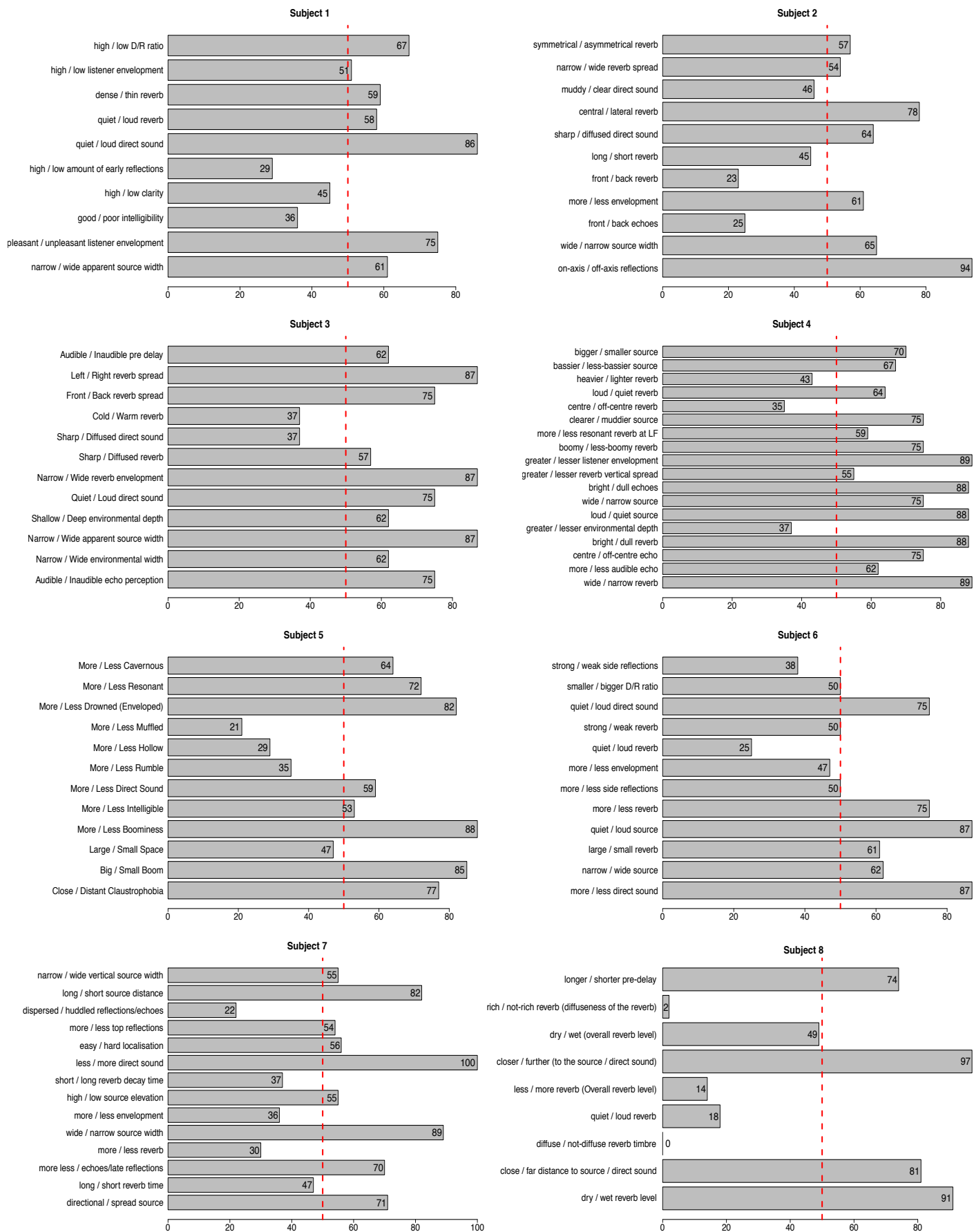

Figure 6. Audibility rating for each subject.

Table 4. Most salient attributes based on audibility index.

| Attribute | No. of occurences | % | Audibility mean | % | Audibility index |
|---|---|---|---|---|---|
| Quiet / Loud - Direct Sound | 6 | 75,00 | 78,75 | 78,75 | 76,88 |
| More / Less - Listener Envelopment | 7 | 87,50 | 63,43 | 63,43 | 75,46 |
| Wide / Narrow - Apparent Source Width | 6 | 75,00 | 66,00 | 66,00 | 70,50 |
| Narrow / Wide - Reverb Spread | 6 | 75,00 | 47,00 | 47,00 | 61,00 |
| Close / Far - Source Distance | 2 | 25,00 | 85,50 | 85,50 | 55,25 |
| Quiet / Loud - Reverb | 5 | 62,50 | 46,05 | 46,05 | 54,28 |
| Central / Lateral - Echoes | 3 | 37,50 | 71,00 | 71,00 | 54,25 |
| Central / Lateral - Reverb | 3 | 37,50 | 63,50 | 63,50 | 50,50 |
| Bright / Dull - Reverb | 1 | 12,50 | 88,00 | 88,00 | 50,25 |
| Bright / Dull - Echoes | 1 | 12,50 | 88,00 | 88,00 | 50,25 |
| Long / Short - Pre-Delay | 2 | 25,00 | 68,00 | 68,00 | 46,50 |
| Audible / Inaudible - Echoes | 3 | 37,50 | 55,33 | 55,33 | 46,42 |
| Boomy / Less-boomy - Reverb | 2 | 25,00 | 63,00 | 63,00 | 44,00 |
| Pleasant / Unpleasant - Listener Envelopment | 1 | 12,50 | 75,00 | 75,00 | 43,75 |
| High / Low - Clarity | 3 | 37,50 | 49,83 | 49,83 | 43,67 |
| High / Low - D/R Ratio | 2 | 25,00 | 60,00 | 60,00 | 42,50 |
| Bright / Dull - Source | 1 | 12,50 | 67,00 | 67,00 | 39,75 |
| Front / Back - Reverb | 2 | 25,00 | 52,00 | 52,00 | 38,50 |
| Narrow / Wide - Environmental Width | 1 | 12,50 | 62,00 | 62,00 | 37,25 |
| Shallow / Deep - Environmental Depth | 2 | 25,00 | 49,50 | 49,50 | 37,25 |
| Large / Small - Space | 1 | 12,50 | 62,00 | 62,00 | 37,25 |
| Clear/ Muddy - Source | 2 | 25,00 | 48,00 | 48,00 | 36,50 |
| Long / Short - Reverb Time | 2 | 25,00 | 43,50 | 43,50 | 34,25 |
| Sharp / Diffused - Direct Sound | 1 | 12,50 | 55,00 | 55,00 | 33,75 |
| Narrow / Wide - Vertical Source Width | 1 | 12,50 | 55,00 | 55,00 | 33,75 |
| More / Less - Vertical Reverb Spread | 1 | 12,50 | 55,00 | 55,00 | 33,75 |
| High / Low - Source Elevation | 1 | 12,50 | 55,00 | 55,00 | 33,75 |
| More / Less - Top Reflections | 1 | 12,50 | 54,00 | 54,00 | 33,25 |
| Cold / Warm - Reverb | 1 | 12,50 | 37,00 | 37,00 | 24,75 |
| More / Less - Rumble | 1 | 12,50 | 35,00 | 35,00 | 23,75 |
| More / Less - Hollow | 1 | 12,50 | 29,00 | 29,00 | 20,75 |
| Front / Back - Echoes | 1 | 12,50 | 25,00 | 25,00 | 18,75 |

## 4. DISCUSSIONS

This section links together the experimental results with the initial aim of the study and proposes set of attribute scales for further research.

### 4.1 Lab Experiment Validity

As discussed above, both the in-situ and lab experiment generated the same clusters in terms of listening positions and orientations. This is important in confirming the validity of the lab elicitation, meaning that in future research a virtual test can be used alone, with a higher efficiency and in a more controlled environment.

The lab experiment was considered to overcome the shortcomings of an in-situ test where the switching of the stimuli cannot be done instantaneously. An initial observation of the elicited constructs showed an increased number of constructs describing timbral differences of different listening positions. It is considered that the ability to quickly switch between stimuli facilitated the perception of timbral differences, which might have been more subtle than spatial ones and therefore more difficult in the in-situ experiment. Furthermore, the audibility-based filtering performed on the lab experiment results extracted a more focused set of perceived attributes. Recognizing the benefit of audibility testing in an elicitation scenario, it is to be noted that it can only be performed in a lab-based framework where back-to back assessment of stimuli is facilitated. Based on the aforementioned aspects, it can be affirmed that the use of a lab elicitation offers certain advantages in terms of robustness, scalability and efficiency while still maintaining the ecological validity of an in-situ test.

### 4.2 Listening Positions and Head Orientations

The clustering showed an interesting pattern in the perception of different positions in a concert hall. Figure 7 provides a visual representation of the clustering. Conditions P1_0°, P1_90° and P1_180° are tightly clustered, meaning a high similarity between them. At this position, it can be suggested that the head orientation does not influence the perception of spatial impression greatly. This might be caused by the level of the direct sound mostly, overpowering many of the other perceptual attributes. The second cluster is formed by P2_0° and P2_180°. Both of these positions are in-line with the central axis of the room, meaning that the overall spatial impression in not greatly changed by a 180° rotation. However, is was observed that differences in the direction of arrival of the early reflections and reverberation (front-back) were elicited. In the third cluster, conditions P2_90°, P3_90° and P4_90°, which are facing 90º from the source can be seen. This is an interesting result, as it suggests that the perception of spatial impression for a 90° head orientation is different from one facing directly towards the source.

P2_90° - which is on the central axis - falls in the same cluster with P3_90° and P4_90° - positioned laterally - suggesting that changes in perceived spatial impression evoked by the head orientation are independent of the lateral positioning of the listener. This opens a new research path in studying the way in which different head orientations influence the perception of spatial impression attributes. Finally, the fourth cluster presents conditions P3_0° and P4_0°, facing towards the source from a lateral listening position. The perceived differences in spatial impression are thought to be due to the asymmetrical arrival of early reflections and reverberant tail as suggested by the elicited constructs. These results are encouraging for further research, looking into the influence of lateral movement on the perception of spatial impression attributes as well as on preference.

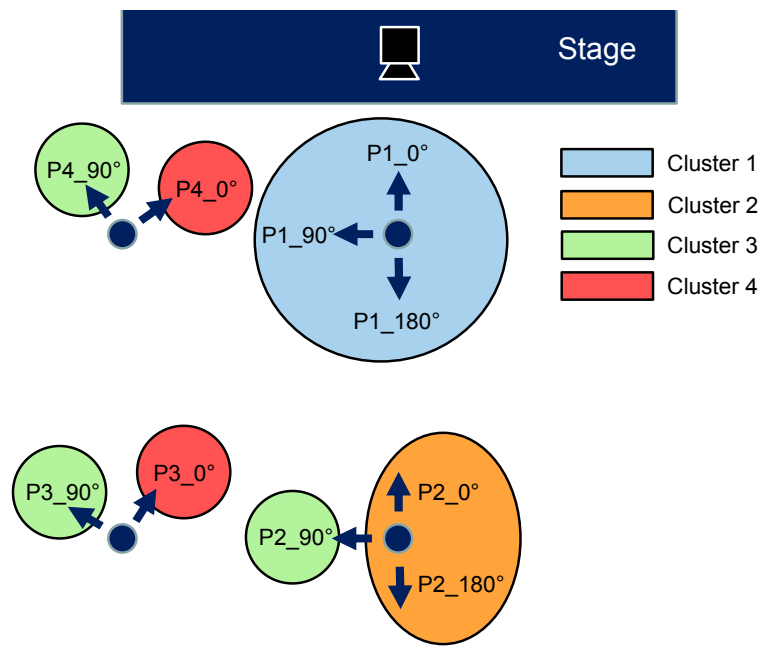

Figure 7. Visual representation of the position clusters

## 4.3 Elicited Constructs

From Table 4, ten attributes with the audibility index greater than 50 were considered to be considerably salient. Table 5 presents proposed labels for these attributes along with their respective definitions in descending order of their audibility index.

Table 5. Salient attributes with their definitions and end-labels

| Attribute | Definition | End labels |
|---|---|---|
| Perceived Source Loudness | The point-source loudness perceived at different locations | Quiet - Loud |
| Listener Envelopment | The feeling of being surrounded by the reverberant field | Less - More |
| Apparent Source Width | The perceived impression of width given by the sound source fused with the early reflections | Narrow - Wide |
| Apparent Reverb Width | The perceived impression of width given by the late reverberant field. | Narrow - Wide |
| Apparent Source Distance | The perceived distance between the listener and the sound source | Close - Far |
| Perceived Reverb Loudness | The reverb loudness perceived at different locations | Quiet - Loud |
| Echo Direction | The perceived direction of arrival for echoes | Central - Lateral |
| Reverb Direction | The perceived direction of arrival for reverberation | Central - Lateral |
| Reverb Brightness | The timbral characteristics of reverberation depending on the level of high frequencies | Bright - Dull |
| Echo Brightness | The timbral characteristics of echoes depending on the level of high frequencies | Bright - Dull |

The attribute with the highest audibility index was Perceived Source Loudness (PSL). This result was expected, as sound pressure level (SPL) depends on the inverse square law. The perception of source distance is also identified as one of the most salient attributes in Table 5, and it is labelled here as Apparent Source Distance (ASD). While PSL and ASD might initially seem to be obvious attributes to be elicited from different physical source-listener distances, the potential dependencies of their perceived magnitudes on the listener's position and head orientation require systematic investigations since physical and perceived quantities may not be consistent. Zahorik and Wightman [38] found that humans tend to overestimate the perceived distance of sound source (i.e., ASD) within about 1m source-listener distance, but underestimate it beyond that. A similar trend was also observed by Kearney et al. [39]. However, the previous research considered sound source and listener positions only in the central axis of the space, with the listener always facing towards the source. Our results indicating PSL and ASD as salient differences perceived across different listener positions and head orientations might be associated with other perceptual cues than just the SPL that depends on the physical source-listener distance. With head oriented towards 90° to the left from the on-axis of the source (e.g., P1-0°), the direct-to-reverberant energy (D/R) ratio, which is widely known as the absolute cue of ASD [40], would be higher for the right ear-input signal than for the left ear one. Furthermore, considering the binaural unmasking phenomenon [41], the source masking of reverberation arriving from the back of the concert hall might be weaker at the 90° orientation compared to 0°, which would also imply that Perceived Reverb Loudness (PRL) would be greater at the contralateral ear than at the ipsilateral ear. This might eventually lead to some differences in the perceived magnitudes of ASD, PSL and

PRL as a function of head orientation even at the same source-listener distance. This hypothesis might also be an explanation for the finding by Li and Peissig [42] that reverberation received by the contralateral ear plays a more important role than that received by the ipsilateral ear on perceived externalisation in the auralisation of a room over headphones. The perceptual mechanism of ASD, PSL and PRL perceptions depending on listener position and head orientation will be quantitatively investigated in our future study.

It is considered that the ASD and PSL attributes might also contribute to the perception of Apparent Source Width (ASW), which was rated as the third most salient attribute in the current study. The basis for this argument is as follows. Lee [18] found that the perceived magnitude of ASW decreased linearly per doubling the distance between 3m, 6m and 12m. It was also reported that the perceived result was not predicted accurately by conventional objective measures for ASW such as IACC Early [43] and Early Lateral Energy Fraction ($Lf$) [4], whereas it had a high positive correlation with Sound Strength ($G$) [44] The importance of $G$ on ASW was also claimed by Marshall and Barron [45], who proposed an objective measure Degree of Source Broadening (DSB) employing both $G$ and $Lf$. However, studies in the context of concert hall acoustics confirmed the IACC Early and $Lf$ as well as [1-IACC E3] (a.k.a. Binaural Quality Index) by Hidaka et al. [9] to be successful measures for evaluating perceptual differences between different concert halls [9, 45, 42, 46]. It should be noted that these studies used source-received distances larger than around 10m, where reflections and reverberation energies are more dominant than the direct sound energy.

From the above, it is hypothesised that the ASW elicited in the present study might have had two perceptual dimensions: one that is related to $G$, which would determine the PSL perception, and the other associated with IACC or/and $Lf$, which would lead to a width perception influenced by the decorrelation and lateral energies of ear signals resulting from the combination of direct sound and early reflections. This is supported by the elicited terms from Subject no. 2, who mentioned both "sharp/focused direct sound" and "wide/narrow source" (Figure 6). From the communication with the subject after the audibility grading experiment, it was identified that he indeed meant two different types of auditory sensations: direct-sound-related and reflection-related ones. As implied from [18] and previous concert hall research mentioned above, ASW would likely be determined predominantly by $G$ when the listener is positioned within the critical distance of a space, whereas IACC or/and $Lf$ might be a better predictor for ASW beyond that point. The critical distance of the concert hall used in our study was around 7m for the given source position, meaning that the two back positions (P2 and P3) were outside the critical distance. Therefore, it can be considered that the balance between the properties of ASW might have varied depending on their position.

Another dominant attribute, Listener Envelopment (LEV) is conventionally defined by the feeling of being surrounded in a reverberant sound field [8, 9]. It is a high-level concept and considered to have multidimensional nature based on the observations from the current results and the literature. As with ASW, conventional objective measures for LEV include a $G$-based one (e.g., Late Lateral Energy Fraction ($LG_{80}^{\infty}$) [8]) and a binaural-decorrelation-based one (e.g., IACC L3 [9]). The present study elicited the attributes labelled as Perceived Reverb Loudness (PRL) and Apparent Reverb Width (ARW), which also seem to be related to $G$ and binaural decorrelation, respectively. In addition, another main attribute elicited, Reverb Direction is also deemed to be associated with both PRL and ARW, which is discussed more below. From this, it is hypothesised that these three reverberation-related attributes are low-level components of LEV.

These attributes have not been reported in previous concert hall studies on spatial impression, which mostly focused on comparing different concert halls in LEV and ASW at similar source-listener distances. Lokki et al. [11] elicited various subjective terms to describe perceived characteristics of different concert halls using a multichannel auralisation technique, and found that "loudness", "distance" as well as "reverberance" were most salient differences perceived *between* concert halls. However, in the current study, PRL, ARW and Reverb Direction were elicited in relation to changes in listener position or/and head orientation *within* a single concert hall. The perception of PRL in this context is particularly interesting since it is usually understood that the physical level of reverberation would vary little across different positions in a space, and yet differences in "perceived" loudness of reverberation were clearly audible. It is not yet possible to establish whether it was a change in listener position, head orientation or both that resulted in a difference in PRL from the result of this elicitation study. This will be investigated further in our future quantitative experiment. However, it is hypothesised tentatively that the PRL perception is mainly due to the binaural unmasking as already described above. Additionally, given the conventional measure of LEV, $LG_{80}^{\infty}$, it seems logical to hypothesise that reverberation arriving laterally would have a greater contribution to the perceptions of PRL and LEV than that from the same or opposition direction of the source.

ARW, on the other hand, is considered to be related to the directional balance of reverberation and the resulting IACC depending on both listener position and head-orientation. For instance, at P1_0° (at the central axis, with the listener's head pointing forward), the left and right ears would receive reverberation from both left and right sides of the concert hall with similar energies and times of arrival. At P1_90°, the reverberation from the back side of the hall would arrive at ears later, but with potential more energy than that from the front of the concert hall, since there is more space for the reverberation to develop in the back. Similarly, at P4_0°, the ears would receive stronger reverberation from the right than from the left. This might eventually result in ear signals with a lower IACC, thus producing a larger ARW. The difference in the time of arrival caused by changes in head orientation and listener position also seems to be the cause for the Echo Direction attribute.

In addition, ARW may initially seem to be a similar concept to Environment Width as a subcomponent of LEV that was proposed by Rumsey [10]. Although Environment Width is also related to reverberant sound field, under his definition, it is about the perception of the "boundaries" of the space in sound reproduction, which requires a sense of "being inside an enclosed space" as a prerequisite. ARW, on the other hand, is quantified by the perceived magnitude of the spread of reverberant sound image, which results from physical causes.

Last but not least, timbral characteristics such as Reverb Brightness and Echo Brightness were also found to be salient attributes. Brightness is widely known to be a property of high frequencies of sound. Hawkes et al. [1] link the term "Brilliance" with the length of reverberation time at high frequencies in the context of audience perception in concert halls. It is interesting that the brightness of late arriving sounds such as reverberation and echo was given a higher audibility index (50.3) than that of source (40). This could be explained as follows. When rotating the head at a given position, the perceived tonal balance of the source would be primarily changed by the head-related transfer function (HRTF) of the given azimuth angle of the source, along with the influence of early reflections that would become stronger at a farther source-distance distance. For late sounds, when the source azimuth is 0° from a given listening position, those arriving from around the front and back of the space may be perceived to be least bright since the masking effect would be the strongest when the masker and maskee are on the same azimuth [48]. However, as the source azimuth becomes more lateral with head rotation, there would be a weaker masking of the late sounds from the front and back of the space (i.e., binaural unmasking), which may increase their perceived brightness.

In addition, it is hypothesised that the brightness of the late sounds may also contribute to the overall clarity of the perceived sound together with the brightness of the early sounds in general. Conventionally, when objectively measuring clarity (e.g., C80), only the energy difference of the early sound to the late sound is considered. However, the current finding suggests that it would be worthwhile to investigate the influence of the brightness of the late sounds on general perceived clarity across multiple listening positions and head orientations.

## 4. CONCLUSIONS

This paper presented a study into the perception of perceived spatial impression in a concert hall, at different listening positions and head orientations. Two complementary experiments were conducted for the elicitation of perceptual attributes; an in-situ experiment which maintained the ecological validity of the actual space and a laboratory experiment which facilitated the elicitation process and addressed some limitations observed in the in-situ experiment. Repertory grid technique (RGT) was used as the elicitation methodology for its robustness and structured approach towards stimuli comparison. An additional audibility test was carried out during the laboratory experiment which facilitated the filtering of salient perceived spatial impression attributes. Verbal protocol analysis was used for the classification of the elicited attributes based on their semantic meaning. Cluster analysis was used for grouping the elicited attributes based on the ratings received in the construct rating stage of RGT, thus facilitating the observation of different perceptual effects. Results showed a good agreement between the in-situ and laboratory experiments. In terms of listening positions and head orientations, the resulting clusters matched exactly between two tests. Ten salient attributes were presented, and future experiments were proposed for in-depth research into the interaction between the attributes and the different listening positions and head orientations. It is believed that the Perceived Source Loudness (PSL) and the Apparent Source Distance (ASD) can affect the more established Apparent Source Width (ASW) attribute. It was also theorised that the Listener Envelopment (LEV) is comprised of two perceptual sub-components: Perceived Reverb Level (PRL) and the Apparent Reverb Width (ARW). It is also believed that the Perceived Reverb Level may be affected by the directionality of the echoes and reverberation. The findings of this study have various useful practical implications, for example on the dynamic rendering of auditory spatial impression in the context of 6DOF XR applications, where the user can virtually navigate through a space or is teleported to various positions. One other application might be in real concert halls, for the prediction of different attributes' (e.g., LEV) variation depending on different seating positions. Future experiments testing these hypotheses will be carried out for a more in-depth understanding of how different spatial impression attributes are perceived in the context of multiple positions and head orientations. Subjective grading tests will be carried out in a controlled environment and will assess the salient spatial impression attributes in a multi-position and multi-head-orientation context.

# Appendices

Appendix 1. Elicited terms along with the RGT ratings from the in-situ test

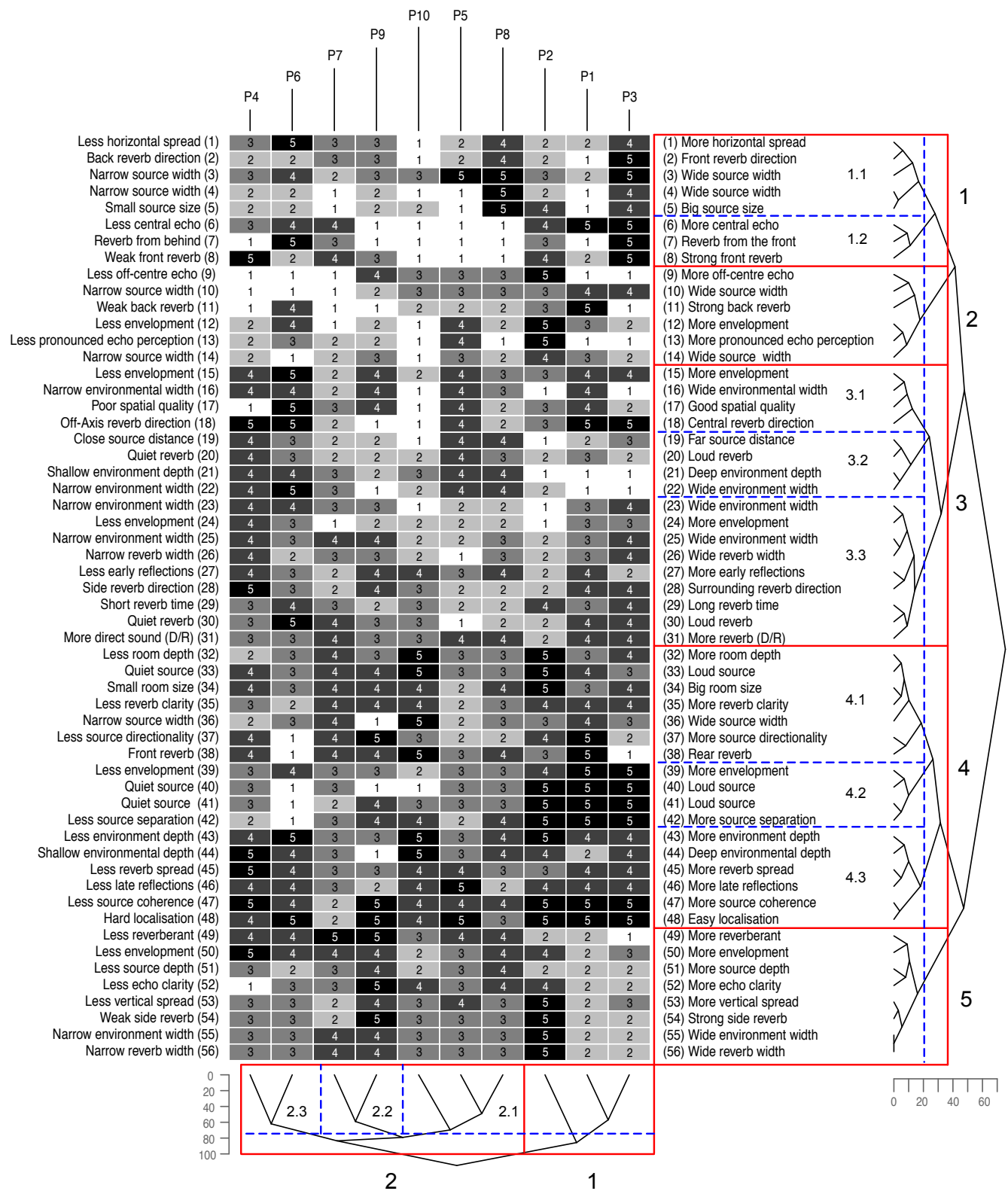

Appendix 2. Elicited terms along with the RGT ratings from the laboratory test.

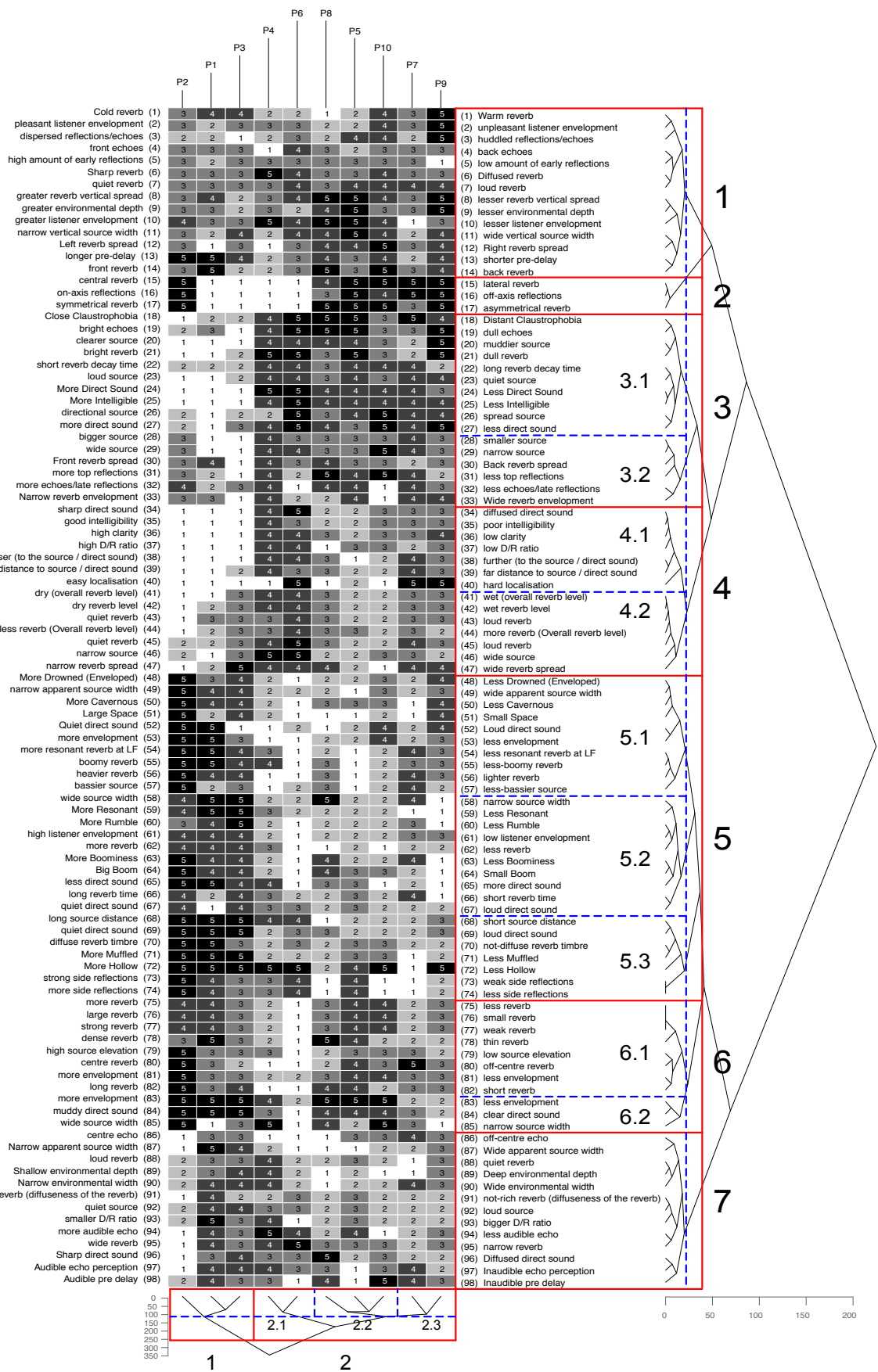